\documentclass[12pt]{article}
\usepackage{epsf,amssymb,amsmath,latexsym}
\usepackage{graphicx}
\usepackage[final]{pdfpages}

\title{Sturm intersection theory for periodic Jacobi matrices \\
and linear Hamiltonian systems}

\author{Hermann Schulz-Baldes
\\
\\
{\small Department Mathematik, Universit\"at
Erlangen-N\"urnberg, Germany}
}

\date{ }

\newtheorem{theo}{Theorem}

\newtheorem{proposi}{Proposition}

\newcommand{\CM}{{\mathbb C}}

\newcommand{\RM}{{\mathbb R}}
\newcommand{\SM}{{\mathbb S}}

\newcommand{\LM}{{\mathbb L}}

\newcommand{\Pp}{{\cal P}}

\newcommand{\Oo}{{\cal O}}
\newcommand{\Tr}{\mbox{\rm Tr}}
\newcommand{\Tt}{{\cal T}}
\newcommand{\Rr}{{\cal R}}
\newcommand{\Vv}{{\cal V}}

\newcommand{\Mm}{{\cal M}}
\newcommand{\Cc}{{\cal C}}
\newcommand{\Jj}{{\cal J}}

\newcommand{\Qq}{{\cal Q}}

\newcommand{\one}{{\bf 1}}

\begin{document}

\maketitle

\begin{abstract}
Sturm-Liouville oscillation theory for periodic Jacobi operators with matrix entries is discussed and illustrated. The proof simplifies and clarifies the use of intersection theory of Bott, Maslov and Conley-Zehnder. It is shown that the eigenvalue problem for linear Hamiltonian systems can be dealt with by the same approach.
\end{abstract}

\vspace{.8cm}

\section{Oscillation theorem for periodic Jacobi matrices}
\label{sec-JMME}

Let $L\geq 1$ and $N\geq 3$ be integers and $\omega\in\SM^1$ or $\omega=0$. A Jacobi matrix with matrix entries is an operator of the form
\begin{equation}
\label{eq-matrix}
H_N({\omega})
\;=\;
\left(
\begin{array}{ccccccc}
V_1       & T_2  &        &        &         & \overline{\omega}\, T^*_1       \\
T_2^*      & V_2    &  T_3  &        &         &        \\
            & T_3^* & V_3    & \ddots &         &        \\
            &        & \ddots & \ddots & \ddots  &        \\
            &        &        & \ddots & V_{N-1} & T_N   \\
\omega\,T_1     &        &        &        & T_N^*  & V_N
\end{array}
\right)
\;,
\end{equation}
where $(V_n)_{n=1,\ldots,N}$ are selfadjoint complex $L\times L$ matrices and 
$(T_n)_{n=1,\ldots,N}$ are invertible complex $L\times L$ matrices. If $\omega=0$, then the Jacobi matrix is said to have Dirichlet boundary conditions (on the left and right edge) and we denote it by $H^{\mbox{\rm\tiny D}}_N$. If $\omega=e^{\imath k}$ for $k\in[0,2\pi)$, the Jacobi matrix is called periodic and will be denoted by $H^{k}_N$. Both $H^{k}_N$ and $H^{\mbox{\rm\tiny D}}_N$  are selfadjoint operators on the Hilbert space $\ell^2(\{1,\ldots,N\},\CM^L)$. Matrices of type \eqref{eq-matrix} appear in a number of applications. In an example coming from solid state physics,  $H_N^{k}$ is the finite volume approximation of a $d$-dimensional tight-binding hopping model with periodic boundary conditions on a discretized cube $N^d$. The fibers are then of dimension $L=N^{d-1}$ and the $V_n$ describe the model within the fibers while  the $T_n$ model the couplings between them. The purpose of this note is to present an algorithm to calculate the eigenvalues of $H_N^{k}$ which extends the well-known Sturm-Liouville oscillation theory for one-dimensional second-order differential equations. The main results are resembled in the following theorem:

\begin{figure}
\begin{center}
\includegraphics[width=0.47\textwidth]{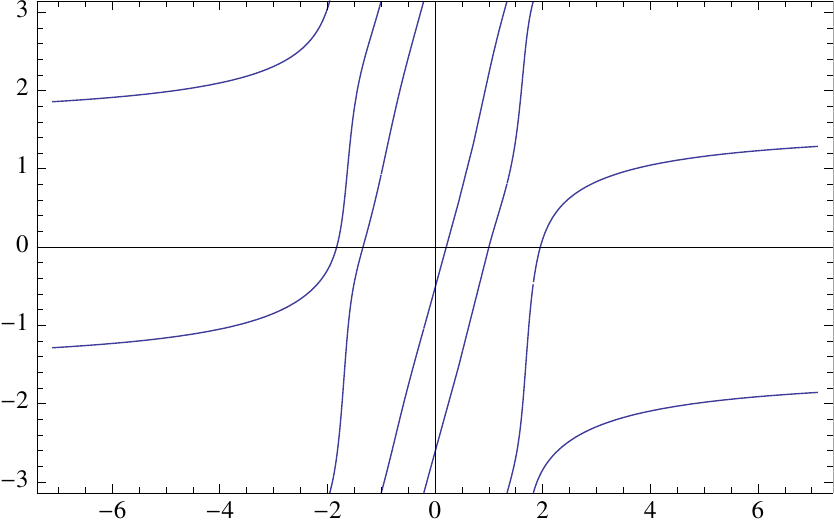}
\includegraphics[width=0.47\textwidth]{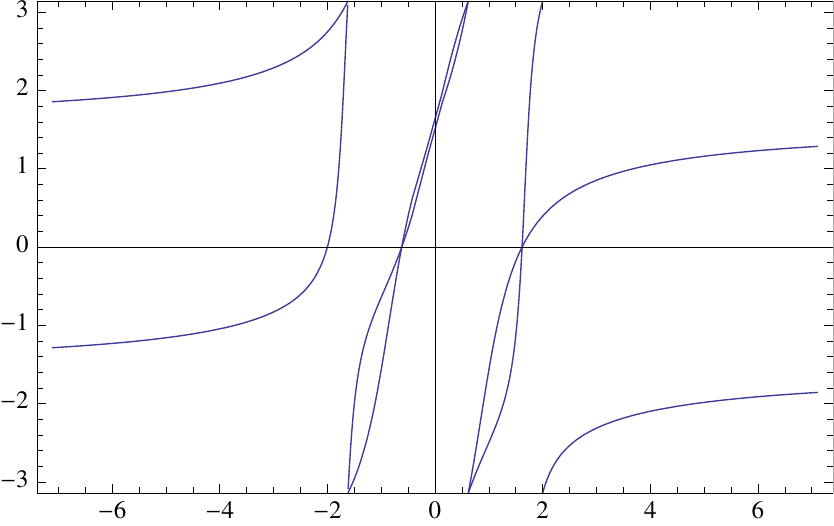}
\caption{\it Plotted are the arguments of the two eigenvalues of $\widehat{U}^{E,k}_N$ as a function of the energy for the Hamiltonian $H^k_N$ with a simple fiber {\rm ($L=1$)} and for $N=5$ with $T_n=1$, $V_n=0$ and $k=\frac{\pi}{3}$ {\rm (}left figure{\rm )} and $k=\pi$ {\rm (}right figure{\rm )}. The eigenvalues are at the intersections with the axis, namely at $-1.83$, $-1.34$, $0.21$, $1.0$ and $1.96$ for $k=\frac{\pi}{3}$, and at $-2.0$, $-0.62$ and $1.62$ for $k=\pi$. One also sees that one argument approaches $\frac{\pi}{2}$ from above as $E\to-\infty$, while another approaches $-\frac{\pi}{2}$ in that limit. For $E\to\infty$ there is a similar behavior. 
}
\end{center}
\end{figure}

\begin{theo} 
\label{theo-osci}
Associated to $H_N^{k}$ there is a real analytic path $E\in\RM\mapsto\widehat{U}^{E,k}_N$ of $2L\times 2L$ unitary matrices such that the following hold:

\vspace{.1cm}

\noindent {\rm (i)} As a function of energy $E$, the eigenvalues of $\widehat{U}^{E,k}_N$ rotate around the unit circle in the positive 

sense and with non-vanishing speed.

\vspace{.1cm}

\noindent {\rm (ii)} As $E\to\pm\infty$, half of the eigenvalues of  $\widehat{U}^{E,k}_N$ converge to $\imath$ and the other half to $-\imath$.

\vspace{.1cm}

\noindent {\rm (iii)} The multiplicity of $E$ as eigenvalue of $H^k_N$ equals the multiplicity of $1$ as eigenvalue of $\widehat{U}^{E,k}_N$.

\end{theo}

The theorem is illustrated by two numerical examples in Figure 1 and 2. Of course, an important point explained below is how to calculate the unitaries $\widehat{U}^{E,k}_N$. Equation \eqref{eq-unitaryMoebius} below shows that $\widehat{U}^{E,k}_N$ can be calculated iteratively using the matrix M\"obius transformation. In the example of a tight-binding Hamiltonian cited above, Theorem~\ref{theo-osci} thus reduces the linear algebra problem by one dimension from $d$ to $d-1$, because the M\"obius transformation involves inverting $L\times L$ matrices. However, this has to be done many times in order to deduce the eigenvalues of $H^k_N$ using Theorem~\ref{theo-osci}. Hence, whether this procedure is of any numerical interest when it comes to calculating the eigenvalues of matrices of type \eqref{eq-matrix} is not clear to the author. On a theoretical level, the arguments below leading to Theorem~\ref{theo-osci} show that $\widehat{U}^{E,k}_N$ naturally describes the intersection theory of two Lagrangian planes, one given by the formal solutions of the Schr\"odinger equation associated to $H^k_N$ and the other by a boundary condition modeling the periodicity.  

\vspace{.2cm}

\begin{figure}
\begin{center}
\includegraphics[width=0.47\textwidth]{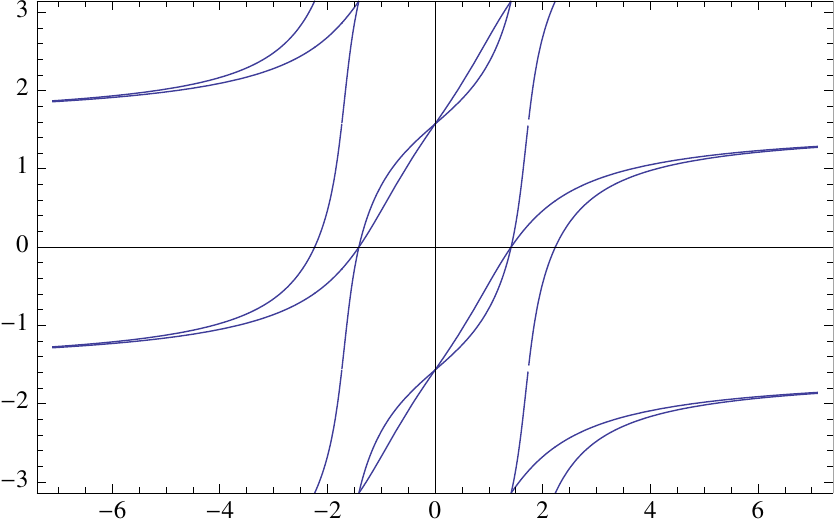}
\includegraphics[width=0.47\textwidth]{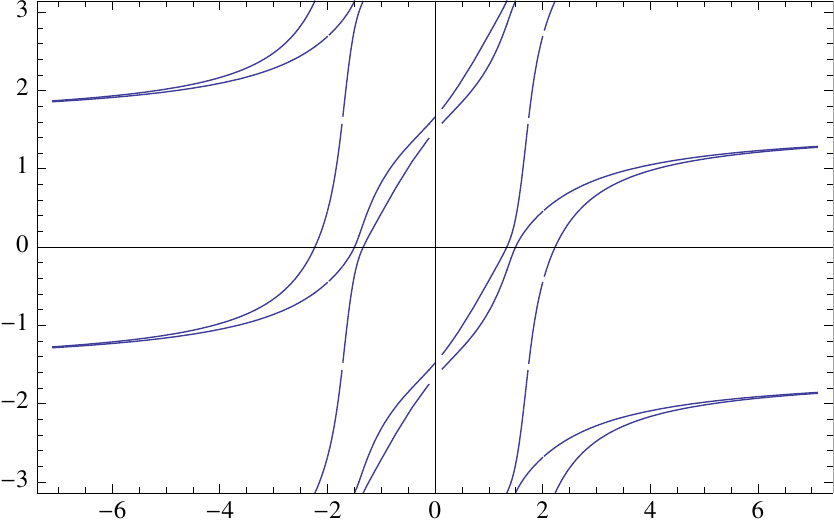}
\caption{\it Plotted are the arguments of the four eigenvalues of $\widehat{U}^{E,k}_N$ as a function of the energy for the Hamiltonian $H^k_N$ with a double fiber {\rm ($L=2$)} and for $N=3$ with 
$T_n=\binom{0 \;1}{1\;0}$, $V_n=\binom{1 \;\;0}{0\;-1}$ and $k=0$ {\rm (}left figure{\rm )} and $k=0.6$ {\rm (}right figure{\rm )}. The $6$ eigenvalues are at the intersections with the axis and the asymptotics for large $E$ are also seen. Unfortunately, the graphics become less trackable for larger $N$ and $L$.
}
\end{center}
\end{figure}

Before the more technical part of the paper, let us review the history of Sturm-Liouville oscillation theory for matrix-valued operators. The first result is due to Lidskii \cite{Lid} who considered Hamiltonian systems and implicitly studied the corresponding operators with Dirichlet boundary conditions. Independently and motivated by the closed geodesic problem,  Bott \cite{Bot} studied matrix-valued periodic Sturm-Liouville operators and proved a version of the above theorem.  To our best knowledge, the case of Jacobi matrices with matrix-valued entries was first considered in our prior work \cite{SB}. It contained only separated left and right boundary conditions and, in particular, the case of  $H^{\mbox{\rm\tiny D}}_N$. This result is recalled in Section~\ref{sec-transfer}. To round things up, Section~\ref{sec-HamiltonianSys} briefly shows how the very same techniques can be applied to Hamiltonian systems and thus provides an alternative proof of the result of Bott \cite{Bot}.

\vspace{.2cm}

The main reason why we believe that the present proof considerably simplifies prior ones is the following: the boundary conditions are (hermitian) Lagrangian planes and these Lagrangian planes can be identified with the unitary matrices (Proposition~\ref{prop-diffeo}). For the spectral problem, one is then led to study the intersection of the formal solutions of the Schr\"odinger equation with the boundary conditions.  The intersection of two Lagrangian planes can easily be read off the spectrum of the associated unitaries (Proposition~\ref{prop-Wronski}). The Bott-Maslov index of a path of Lagrangian planes counts the number of these intersections and can thus simply be defined by looking at the spectrum of the associated unitaries (Section~\ref{sec-index}). The path relevant for the eigenvalue problem of $H^{\mbox{\rm\tiny D}}_N$ is given by the energy dependence of Lagrangian planes naturally associated to the solutions of the Schr\"odinger equation $H^{\mbox{\rm\tiny D}}_N\phi=E\phi$. For the study of the periodic operators $H^k_N$, one also needs the Bott or Conley-Zehnder index of a path of hermitian symplectic matrices. But using a variant of the graph of each such matrix (which is also Lagrangian) this index is just a special case of the Bott-Maslov index in doubled dimension (Section~\ref{sec-index}).

\vspace{.2cm}

Many of the facts about intersection theory of Lagrangian planes and symplectic paths are well-known (even though in general stated for real symplectic and not hermitian symplectic structure), but the consequent use of the identification of Lagrangian planes with unitaries leads to a particularly intelligible and compact presentation below. The first such intersection theory seems to be due to Bott \cite{Bot} and was developed precisely for the eigenvalue problem of matrix-valued Sturm-Liouville operators, see Section~\ref{sec-HamiltonianSys}.  Nevertheless, the theory is nowadays most often associated with Maslov, due to his contributions \cite{Mas}. Conley and Zehnder considered paths of symplectic matrices and defined their index \cite{CZ}, but this could also be considered a further development of Bott's ideas. A spectral flow definition of the index is due to Robbin and Salomon \cite{RS}. Extensions, further geometric applications as well as a nice review on the indices as well as eigenvalue problems are given by Long \cite{Lon}. Another recent review on various approaches to the above indices and links between them is \cite{Gos}. Sections~\ref{sec-symplectic} and \ref{sec-index} resembles those results relevant for the proof of Theorem~\ref{theo-osci} without further references, but also no claim of novelty.

\vspace{.2cm}

\noindent {\bf Acknowledgement:} The author thanks A. Knauf for discussions and a referee for a number of useful suggestions which allowed to improve the paper. This work was supported by the DFG.

\section{Hermitian symplectic planes and matrices}
\label{sec-symplectic}

Let the symplectic structure be given in its standard form:
$$
\Jj\;=\;
\left(
\begin{array}{cc}
0 & -\one_{L} \\
\one_L & 0
\end{array}
\right)
\;.
$$
We recall that (hermitian symplectic) Lagrangian planes in $\CM^{2L}$ are $L$-dimensional planes on which the symplectic form $\Jj$ vanishes. Such planes can be described by $L$ linearly dependent vectors $\phi(1),\ldots,\phi(L)\in\CM^{2L}$, which we regroup to an $2L\times L$ matrix $\Phi=(\phi(1),\ldots,\phi(L))$ of rank $L$. For a Lagrangian plane this matrix then satisfies $\Phi^*\Jj\Phi=0$ and we then call $\Phi$ also a Lagrangian frame.  Actually, several frames ${\Phi}$ describe the same plane which is hence specified by an equivalence class $[{\Phi}]_\sim$ with respect to the relation ${\Phi}\sim\Phi'\Longleftrightarrow {\Phi}=\Phi'c$ for some $c\in\mbox{\rm Gl}(L,\CM)$. The Lagrangian Grassmannian $\LM_L$ is the set of all equivalence classes of Lagrangian frames. If a plane has a real representative $\Phi$, then it belongs to the real Lagrangian Grassmannian $\LM_L^\RM$ usually considered in symplectic geometry \cite{Arn}. Even though well-known \cite{Bot}, we recall the short argument leading to the following result  for the convenience of the reader because it is central to the approach below. 

\begin{proposi}
\label{prop-diffeo} 
As real analytic manifold, the Lagrangian Grassmannian $\LM_L$ is diffeomorphic to the unitary group $\mbox{\rm U}(L)$ via the stereographic projection $\Pi:\LM_L\to \mbox{\rm U}(L)$ defined by
$$
{\Pi}\bigl([{\Phi}]_\sim\bigr)
\;=\;
\left[\binom{\one_{L}}{\imath\,\one_{L}}^*{\Phi}\right]\,
\left[\binom{\one_{L}}{-\imath\,\one_{L}}^*{\Phi}\right]^{-1}\;.
$$
This map also establishes a diffeomorphism from $\LM_L^\RM$ to the symmetric unitaries in $\mbox{\rm U}(L)$.
\end{proposi}

\noindent {\bf Proof.} 
Let $\Phi=\binom{a}{b}$ be Lagrangian where $a$ and $b$ are $L\times L$ matrices satisfying
$a^*b=b^*a$. Then
\begin{eqnarray*}
L
& = &
\mbox{rank}(\Phi)
\;=\;
\mbox{rank}(\Phi^*\Phi)
\;=\;
\mbox{rank}(a^*a+b^*b)
\nonumber
\\
& = &
\mbox{rank}\bigl((a+\imath \,b)^*(a+\imath \,b) \bigr)
\;=\;
\mbox{rank}(a+\imath \,b)
\;=\;
\mbox{rank}(a-\imath \,b)
\;.
\nonumber
\end{eqnarray*}
It follows that the inverse in ${\Pi}\bigl([{\Phi}]_\sim\bigr)=(a-\imath \,b)(a+\imath \,b)^{-1}$ exists and thus that $\Pi$ is well-defined. Setting $\alpha=a-\imath \,b$ and $\beta=a+\imath \,b$, one has $\alpha^*\alpha=\beta^*\beta$ and ${\Pi}\bigl([{\Phi}]_\sim\bigr)=\alpha\beta^{-1}$. Therefore 
$\Pi([\Phi]_\sim)^*\Pi([\Phi]_\sim)=(\beta^*)^{-1}\alpha^*\alpha\,\beta^{-1}=\one_L$ so that $\Pi([\Phi]_\sim)$ is unitary. Further note that the definition of $\Pi([\Phi]_\sim)$ is indeed independent of the choice of the representative for $[\Phi]_\sim$. Moreover, one can directly check that the inverse of $\Pi$ is given by
\begin{equation}
\label{eq-Piinv}
\Pi^{-1}(U)
\;=\;
\left[
\left(
\begin{array}{c} (U+\one_L) \\  \imath\,(U-\one_L)
\end{array}\right)
\right]_\sim
\;.
\end{equation}
This is clearly real analytic and thus completes the proof of the first statement. It is straightforward to implement the real symmetry, see \cite{SB} where also a symplectic symmetry of the plane is dealt with.
\hfill $\Box$

\vspace{.2cm}

The dimension of the intersection of two Lagrangian planes can be conveniently read off from the spectral theory of the associated unitaries, as shows the next proposition.

\begin{proposi}
\label{prop-Wronski} 
Let ${\Phi}$ and ${\Psi}$ be Lagrangian frames w.r.t. to ${\Jj}$, and let $U=\Pi([\Phi]_\sim)$ and $V=\Pi([\Psi]_\sim)$. Then
$$
\dim\bigl({\Phi}\;\CM^{L}\;\cap\;{\Psi}\;\CM^{L}\bigr)
\;=\;
\dim\bigl(\mbox{\rm Ker}({\Phi}^*{\Jj}\,{\Psi})\,\bigr)
\;=\;
\dim\bigl(\,
\mbox{\rm Ker}(V^*U-\one_L)\,\bigr)
\;.
$$
For the first equality to hold, one only needs $\Phi$ to be Lagrangian.
\end{proposi}

\noindent {\bf Proof.} Let us begin with the inequality $\leq$ of the first equality. Suppose there are two $L\times p$ matrices $v,w$ of rank $p$ such that $\Phi v=\Psi w$. Then $\Phi^*\Jj\Psi w=\Phi^*\Jj\Phi v=0$ so that the kernel of $\Phi^*\Jj\Psi$ is at least of dimension $p$. Inversely, given a $L\times p$ matrix $w$ of rank $p$ such that $\Phi^*\Jj\Psi w=0$, we deduce that $(\Jj\Phi)^*\Psi w=0$. As the column vectors of $\Phi$ and $\Jj\Phi$ are orthogonal and span $\CM^{2L}$, it follows that the column vectors of $\Psi w$ lie in the span of $\Phi$, that is, there exists an $L\times p$ matrix $v$ of rank $p$ such that $\Psi w=\Phi v$. This shows the other inequality and hence proves the first equality of the lemma. For the second,  we first note that the dimension of the kernel of $\Phi^*\Jj\Psi$ does not depend on the choice of the representative. We use the representative of $[\Phi]_\sim$ given in \eqref{eq-Piinv} and a similar one for $[\Psi]_\sim$ in terms of $V$.  But a short calculation then shows that $\Phi^*\Jj\Psi=2\imath\,U^*(U-V)$ which implies the second equality.
\hfill $\Box$

\vspace{.2cm}

Lagrangian planes are mapped to Lagrangian planes by matrices in the hermitian symplectic group  defined by
$$
\mbox{\rm HS}(2L,\CM)
\;=\;
\left\{ \Tt\in\mbox{\rm Mat}(2L\times 2L,\CM)\;\left|
\;\Tt^*\Jj\Tt=\Jj\right.\right\}
\;.
$$
In the literature on Krein spaces, the hermitian symplectic matrices are also called $\Jj$-unitaries \cite{Kre}. If all entries of $\Tt\in\mbox{\rm HS}(2L,\CM)$ are real, then $\Tt$ is in the symplectic group $\mbox{\rm SP}(2L,\RM) $ which acts on real Lagrangian planes. By conjugation with the Cayley transform
$$
{\Cc}
\;=\;
\frac{1}{\sqrt{2}}\;
\left(
\begin{array}{cc}
\one_{L} & -\imath\,\one_{L} \\
\one_{L} & \imath\,\one_{L} 
\end{array}
\right)
\;,
$$
the hermitian symplectic group is isomorphic with the generalized Lorentz group $\mbox{\rm U}(L,L) $ of signature $(L,L)$ conserving the quadratic form $\binom{1\;\,\;0}{0\,-1}$, namely $\mbox{\rm U}(L,L)=\Cc\, \mbox{\rm HS}(2L,\CM)\,\Cc^* $ given explicitly by 
\begin{equation}
\label{eq-ULL}
\mbox{\rm U}(L,L)
\;=\;
\left\{
\left.\,\left(
\begin{array}{cc}
A & B \\  C & D
\end{array}
\right)\;
\right|
\;
A^*A-C^*C=\one_L\;,\;
D^*D-B^*B=\one_L\;,\;
A^*B=C^*D\;
\right\}
\;.
\end{equation}
As already stated, the Hermitian symplectic group acts on the Lagrangian Grassmannian through the map $(\Tt,[\Phi]_\sim])\in\mbox{\rm HS}(2L,\CM)\times\LM_L\mapsto [\Tt\Phi]_\sim\in\LM_L$. We also denote it simply by $\Tt[\Phi]_\sim=[\Tt\Phi]_\sim$. Under the stereographic projection $\Pi$ this action becomes the action of the Lorentz group  $\mbox{\rm U}(L,L) $ via matrix M\"obius transformation on the unitary group:
\begin{equation}
\label{eq-Moeb}
\Pi\bigl([\Tt\Phi]_\sim\bigr)
\;=\;
\Cc\Tt\Cc^*\cdot\Pi\bigl([\Phi]_\sim\bigr)
\;,
\end{equation}
where the dot denotes
$$
\left(
\begin{array}{cc}
A & B \\
C & D
\end{array}
\right)
\cdot U
\;=\;
(AU+B)(CU+D)^{-1}
\;.
$$
This formula also defines an action of $\mbox{\rm U}(L,L) $ on the Siegel disc of matrices satisfying $U^*U<\one$ ({\it e.g.} \cite{FF}), but here we use the action on the maximal boundary of the Siegel disc seen as a stratified space (for further details, see \cite{SB}).

\vspace{.2cm}

Next let us recall the well-known fact that the graph of a hermitian symplectic matrix $\Tt$ is Lagrangian w.r.t. the form $\mbox{\rm diag}(\Jj,-\Jj)$. This can be readily checked by representing the graph by a $2L$-dimensional frame $\binom{\one}{\Tt}$ in $\CM^{4L}$. Another way to state that is that $\binom{\one}{\Tt}=\binom{\one\;\;0}{0\;\;\Tt}\binom{\one}{\one}$ is built from a fixed Lagrangian frame $\binom{\one}{\one}$ w.r.t. $\mbox{\rm diag}(\Jj,-\Jj)$ distorted  by the matrix $\binom{\one\;\;0}{0\;\;\Tt}$ conserving the form $\mbox{\rm diag}(\Jj,-\Jj)$. This matrix $\mbox{\rm diag}(\Jj,-\Jj)$ also conserves the form $\mbox{\rm diag}(\Jj,\Jj)$ and thus, given any Lagrangian plane $\Psi$ w.r.t. this form $\mbox{\rm diag}(\Jj,\Jj)$, one obtains a Lagrangian frame $\binom{\one\;\;0}{0\;\;\Tt}\Psi$ w.r.t. $\mbox{\rm diag}(\Jj,\Jj)$. This is the idea behind the following construction, except that we work with a different representation and a particular $\Psi$ adapted to our purposes below.

\vspace{.2cm}

Hence let us associate to $\Tt\in \mbox{\rm HS}(2L,\CM)$ the $4L\times 4L$ matrix $\widehat{\Tt}=\one_L\widehat{\oplus}\Tt$ where the $\widehat{\oplus}$ denotes the symplectic checker board sum given by
\begin{equation}
\label{eq-expanding}
\left(
\begin{array}{cc}
A & B \\
C & D
\end{array}
\right)
\,\widehat{\oplus}\,
\left(
\begin{array}{cc}
A' & B' \\
C' & D'
\end{array}
\right)
\;=\;
\left(
\begin{array}{cccc}
A & 0 & B & 0 \\
0 & A' & 0 & B' \\
C & 0 & D & 0 \\
0 & C' & 0 & D'
\end{array}
\right)
\;.
\end{equation}
This matrix is in the group $\mbox{\rm HS}(4L,\CM)$ of matrices conserving the symplectic form
$\widehat{\Jj}={\Jj}\,\widehat{\oplus}\,\Jj$. We denote the stereographic projection from $\LM_{2L}$ to $\mbox{\rm U}(2L)$ by $\widehat{\Pi}$ and the $4L\times 4L$ Cayley transform by
$\widehat{\Cc}=\Cc\,\widehat{\oplus}\,\Cc$. Then $\widehat{\Cc}\, \mbox{\rm HS}(4L,\CM)\,\widehat{\Cc}^*$ is the generalized Lorentz group of signature $(2L,2L)$. Furthermore, one can also define the symplectic sum of frames by 
\begin{equation}
\label{eq-Phisum}
\left(
\begin{array}{cc}
a \\
b
\end{array}
\right)
\,\widehat{\oplus}\,
\left(
\begin{array}{cc}
a' \\
b'
\end{array}
\right)
\;=\;
\left(
\begin{array}{cc}
a & 0  \\
0 & a' \\
b & 0  \\
0 &  b'
\end{array}
\right)
\;.\,
\qquad
a,b,a',b'\in\mbox{\rm Mat}(L\times L,\CM)\;.
\end{equation}
If $\Phi$ and $\Psi$ are $\Jj$-Lagrangian, then $\Phi\,\widehat{\oplus}\,\Psi$ is $\widehat{\Jj}$-Lagrangian. We will use a particular $\widehat{\Jj}$-Lagrangian which is not of this form:
\begin{equation}
\label{eq-Phiexpand}
\widehat{\Psi}_0
\;=\;
\left(
\begin{array}{cc}
0 & \one_L \\
\one_{L} & 0 \\
\one_{L} & 0  \\ 
0 & \one_{L} \end{array}
\right)
\;.
\end{equation}
One has $\widehat{\Psi}_0^*\,\widehat{\Jj}\,\widehat{\Psi}_0=0$ so that $\widehat{\Psi}_0$ defines a class in the Lagrangian Grassmannian $\LM_{2L}$. Note that in our notations the hat always designates objects with $L$ replaced by $2L$, but, moreover, $\widehat{\Tt}$ is a particular matrix in $\mbox{\rm HS}(4L,\CM)$ associated to a given $\Tt\in\mbox{\rm HS}(2L,\CM)$. Now as $\widehat{\Psi}_0$ is Lagrangian, so is $\widehat{\Tt}\,\widehat{\Psi}_0$ and we can therefore associated a unitary $\widehat{\Pi}([\widehat{\Tt}\,\widehat{\Psi}_0]_\sim)$ to it. A modification of this unitary will turn out to be particularly useful for the spectral analysis of $\Tt\in\mbox{\rm HS}(2L,\CM)$ (see Proposition~\ref{prop-TtoU} below).

\begin{proposi}
\label{prop-TtoU0} To a given $\Tt\in\mbox{\rm HS}(2L,\CM)$ let us associate a unitary
\begin{equation}
\label{eq-Uhatdef}
\widehat{U}
\;=\;
\left(
\begin{array}{cc}
0 & \imath\,\one_{2L} \\
\imath\,\one_{2L} & 0
\end{array}
\right)
\,\Pi([\widehat{\Tt}\,\widehat{\Psi}_0]_\sim)
\;\in\;
\mbox{\rm U}(2L)
\;.
\end{equation} 
Then 
$$
\widehat{U}
\;=\;
\left(
\begin{array}{cc}
A-BD^{-1}C & \imath\,BD^{-1} \\
\imath \,D^{-1}C & D^{-1} 
\end{array}
\right)
\;,
$$
where the matrices $A,B,C$ and $D$ are given by 
\begin{equation}
\label{eq-CTCcoef}
\Cc\,\Tt\,\Cc^*
\;=\;
\left(
\begin{array}{cc}
A & B \\
C & D
\end{array}
\right)
\;.
\end{equation}
The map $\Tt\in\mbox{\rm HS}(2L,\CM)\mapsto \widehat{U}\in\mbox{\rm U}(2L)$
is a dense embedding  with image
\begin{equation}
\label{eq-Uhatcoeff}
\left\{
\left.\,\left(
\begin{array}{cc}
\alpha & \beta \\ \gamma & \delta
\end{array}
\right)\;\in\;\mbox{\rm U}(2L)\;
\right|
\;
\alpha,\delta\in\mbox{\rm Gl}(L,\CM)
\;
\right\}
\;.
\end{equation}
Thus $\mbox{\rm U}(2L)$ is a compactification of $\mbox{\rm HS}(2L,\CM)$.  If $\Tt\in\mbox{\rm SP}(2L,\RM)$, then $\widehat{U}$ is symmetric. The group $\mbox{\rm SP}(2L,\RM)$ is embedded in the compact space of symmetric unitaries in $\mbox{\rm U}(2L)$.
\end{proposi}

\noindent {\bf Proof.} 
First of all, ${\Cc}\,{\Tt}\,{\Cc}^*$ is in the generalized Lorentz group $\mbox{\rm U}(L,L)$ and thus by \eqref{eq-ULL} one has $D^*D\geq \one_{L}$ so that $D$ is invertible (similarly $A$ is invertible). Now by \eqref{eq-Moeb},
$$
\widehat{\Cc}\,\widehat{\Tt}\,\widehat{\Cc}^*
\cdot
\Pi([\Psi_0]_\sim)
\;=\;
\left(
\begin{array}{cccc}
\one_L & 0 & 0 & 0 \\
0 & A & 0 & B \\
0 & 0 & \one_L & 0 \\
0 & C & 0 & D
\end{array}
\right)
\cdot
\left(
\begin{array}{cc}
0 & -\imath\,\one_L \\
-\imath\,\one_L & 0 
\end{array}
\right)
\;,
$$
so that writing out the M\"obius transformation, one gets
\begin{eqnarray*}
\widehat{\Cc}\,\widehat{\Tt}\,\widehat{\Cc}^*
\cdot
\Pi([\Psi_0]_\sim)
& = & 
\left(
\begin{array}{cc}
0 & -\imath\,\one_L \\
-\imath\,A & B 
\end{array}
\right)
\left(
\begin{array}{cc}
\one_L & 0  \\
-\imath\,C & D 
\end{array}
\right)^{-1}
\\
& = &
\left(
\begin{array}{cc}
0 & -\imath\,\one_L \\
-\imath\,A & B 
\end{array}
\right)
\left(
\begin{array}{cc}
\one_L & 0  \\
\imath\,D^{-1}C & D^{-1} 
\end{array}
\right)\;.
\end{eqnarray*}
From this the formula for $\widehat{U}$ follows and clearly its lower left entry, the matrix denoted by $\delta$ in \eqref{eq-Uhatcoeff}, is invertible. Moreover, by unitarity, $\beta^*\beta+\delta^*\delta=\one_L$ so that $\|\beta\|<1$, which when combined with $\alpha\alpha^*+\beta\beta^*=\one_L$ implies that also $\alpha$ is invertible.

\vspace{.2cm}

Finally let us show that the map $\Tt\in\mbox{\rm HS}(2L,\CM)\mapsto \widehat{U}\in\mbox{\rm U}(2L)$ is surjective onto the set \eqref{eq-Uhatcoeff}. Indeed, given an element of this set, it is natural to set $D=\delta^{-1}$, $B=-\imath\beta\delta^{-1}$, $C=-\imath\delta^{-1}\gamma$ and $A=\alpha-\beta\delta^{-1}\gamma$. With some care one then checks that the equations in \eqref{eq-ULL} indeed hold.
\hfill $\Box$

\vspace{.2cm}

Of course, the entries of $\Cc\Tt\Cc^*$ can be read off from those of $\Tt$:
\begin{equation}
\label{eq-linkC}
\Cc\,\left(
\begin{array}{cc}
A & B\\ C & D
\end{array}
\right)
\,\Cc^*
\;=\;
\frac{1}{2}\;
\left(
\begin{array}{cc}
(A+D)+\imath (B-C) & (A-D)-\imath (B+C)  \\ (A-D)+\imath (B+C)  & (A+D)-\imath (B-C) 
\end{array}
\right)
\;.
\end{equation}
Hence $\widehat{U}$ can also be written out in terms of the entries of $\Tt$.
The following result justifies the above construction and, in particular, the choice of $\widehat{\Psi}_0$.

\begin{proposi}
\label{prop-TtoU}
Let $\Tt$ and $\widehat{U}$ be as in {\rm Proposition \ref{prop-TtoU0}}. For $k\in[0,2\pi)$ introduce the unitary
$$
\widehat{U}^k
\;=\;
\left(
\begin{array}{cc}
e^{-\imath k}\,\one_L & 0 \\
0 & e^{\imath k}\,\one_L 
\end{array}
\right)
\;\widehat{U}
\;.
$$
Then
$$
\mbox{\rm geometric multiplicity of }e^{\imath k}\;\mbox{\rm as eigenvalue of }\Tt
\;=\;
\mbox{\rm multiplicity of }1\;\mbox{\rm as eigenvalue of }\widehat{U}^k\;.
$$
\end{proposi}

\noindent {\bf Proof.} Let us introduce
\begin{equation}
\label{eq-Psik}
\widehat{\Psi}_k
\;=\;
\left(
\begin{array}{cc}
0 &\one_L  \\
e^{\imath k}\,\one_{L} & 0 \\
\one_{L}  & 0 \\ 
0 & e^{\imath k}\,\one_{L} \end{array}
\right)
\;.
\end{equation}
This is a Lagrangian frame w.r.t. $\widehat{\Jj}$. Suppose that this frame and the Lagrangian frame $\widehat{\Tt}\,\widehat{\Psi}_0$ have a non-trivial intersection. This means that there are vectors
$v,w,v',w'\in\CM^L$ such that such $\widehat{\Psi}_k\binom{v}{w}=\widehat{\Tt}\,\widehat{\Psi}_0\binom{v'}{w'}$. The first and third line of this vector equality imply $w=w'$ and $v=v'$, the other two that $\Tt\binom{v}{w}=e^{\imath k}\binom{v}{w}$. This shows 
$$
\mbox{\rm geometric multiplicity of }e^{\imath k}\;\mbox{\rm as eigenvalue of }\Tt
\;=\;
\dim\bigl(\,\widehat{\Tt}\,\widehat{\Psi}_0\,\CM^{2L}\,\cap\,
\widehat{\Psi}_k\,\CM^{2L}\bigr)
\;.
$$
But now Proposition~\ref{prop-Wronski} can be applied to calculate the r.h.s.. As
$$
\widehat{\Pi}\bigl([\widehat{\Psi}_k]_\sim\bigr)
\;=\;
\left(
\begin{array}{cc}
0 & -\imath\, e^{-\imath k}\,\one_{L} \\
-\imath\, e^{\imath k}\,\one_{L} & 0 
\end{array}
\right)
\;,
$$
Proposition~\ref{prop-TtoU0}  completes the proof.
\hfill $\Box$

\section{Intersection indices}
\label{sec-index}

Let us fix a Lagrangian plane $[\Psi]_\sim\in\LM_L$ and define the associated singular cycle $\LM_L^\Psi$ as the stratified space 
$$
\LM_L^\Psi\;=\;
\bigcup_{l=1,\ldots,L}\;
\LM_L^{\Psi,l}
\;,
\qquad
\;\;\;\;\LM_L^{\Psi,l}
\;=\;
\left\{
[\Phi]_\sim\in\LM_L\;\left|
\;
\dim\bigl(\Phi\,\CM^L\cap\Psi\,\CM^L\bigr)=l\,
\right.\right\}
\;.
$$
Under the stereographic projection one gets according to Proposition~\ref{prop-Wronski}
\begin{equation}
\label{eq-PiSing}
\Pi(\LM_L^\Psi)\;=\;
\bigcup_{l=1,\ldots,L}\;
\left\{\;U\in \mbox{\rm U}(L)
\;\left|\;
\dim\,\mbox{\rm ker}\bigl(\Pi([\Psi]_\sim)^*U-\one_L \bigr)=l\,
\right.\right\}
\;.
\end{equation}
Let now $\gamma=(\gamma^E)_{E\in[E_0,E_1)}$ be a (continuous) 
path in $\LM_L$ for which the number of intersections  $\{E\in
[E_0,E_1)\;|\;\gamma^E\in\LM_L^{\Psi}  \}$ is finite and does not contain the initial point $E_0$. We explain below that these transversality and boundary conditions can be considerably relaxed in a straightforward manner. The index of interest here counts the number of intersections of the path with the singular cycle $\LM_L^{\Psi}$ weighted by the orientation of the intersections. According to \eqref{eq-PiSing}, these intersection can be conveniently analyzed by the spectral flow of the unitaries
$$
U^E\;=\;
\Pi([\Psi]_\sim)^*\,
\Pi(\gamma^E)
\;.
$$
At an intersection $\gamma^E\in\LM_L^{\Psi,l}$, let $e^{\imath\theta_1^{E'}},\ldots,e^{\imath\theta_l^{E'}}$ be those eigenvalues of the unitary $U^{E'}$ which are all equal to $1$ at $E'=E$. We call the $\theta_k^{E'}\in[-\pi,\pi)$ also the eigenphases of $U^{E'}$. Choose $\epsilon,\delta>0$ such that $\theta_k^{E'}\in[-\delta,\delta]$ for $k=1,\ldots,l$ and $E'\in [E-\epsilon,E+\epsilon]$ and that there are no other eigenphases in $[-\delta,\delta]$ for $E'\neq E$ and finally $\theta_k^{E'}\neq 0$ for those parameters. Let $n_-$ and $n_+$ be the number of those of the $l$ eigenphases less than $0$ respectively before and after the intersection, and similarly let $p_-$ and $p_+$ be the number of eigenphases larger than $0$ before and after the intersection. Then the signature of $\gamma^E$ is defined by
\begin{equation}
\label{eq-signature}
\mbox{\rm sgn}(\gamma^E)
\;=\;
\frac{1}{2}\;(p_+-n_+-p_-+n_-)
\;=\;
l-n_+-p_-
\;.
\end{equation}
Note that $-l\leq\mbox{\rm sgn}(\gamma^E)\leq l$ and that $\mbox{\rm sgn}(\gamma^E)$ is the effective number of eigenvalues that have crossed $1$ in the counter-clock sense. Furthermore the signature is stable under perturbations of the path in the following sense: if an intersection by $\LM_L^{\Psi,l}$ is resolved by a perturbation into a series of intersections by lower strata, then the sum of their signatures is equal to $\mbox{\rm sgn}(\gamma^E)$. Finally let us remark that, if the phases are differentiable and $\partial_E \theta_k^E\neq 0$ for $k=1,\ldots,l$, then $\mbox{\rm sgn}(\gamma^E)$ is equal to the sum of the $l$ signs $\mbox{\rm sgn}(\partial_E \theta_k^E)$, $k=1,\ldots,l$. Yet another equivalent way to calculate  $\mbox{\rm sgn}(\gamma^E)$ is as the signature of $\frac{1}{\imath}(U^E)^*\partial_EU^E$ seen as quadratic from on the eigenspace of $U^E$ to the eigenvalue $1$ (again under the hypothesis that the form is non-degenerate). Now the intersection number or index of the path $\gamma$ w.r.t. the singular cycle $\LM_L^{\Psi}$ is defined by
\begin{equation}
\label{eq-intersec}
\mbox{\rm I}(\gamma,{\Psi})
\;=\;
\sum_{\gamma^E\,\in\,\LM_L^{\Psi}}
\;\mbox{\rm sgn}(\gamma^E)
\;.
\end{equation}
If the initial point $E_0$ is on the singular cycle, but say the speeds of the eigenvalues passing through $1$ are non-vanishing, the index can still be defined. In order to conserve a concatenation property, we include only the initial point $E_0$ in \eqref{eq-intersec} and not the final point $E_1$.  Furthermore, if a path $\gamma$ is such that it stays on the the singular cycle for an interval of parameters $E$, but it is clearly distinguishable how many eigenvalues pass through $1$ in the process, then the index can be defined as well. We do not write out formal, but obvious definitions in these cases and restrict ourself for sake of simplicity to the transversal case with non-singular end points. It is obvious from the definition that $\mbox{\rm I}(\gamma,{\Psi})$ is a homotopy invariant under homotopies keeping the end points of $\gamma$ fixed.

\vspace{.2cm}

If the path $\gamma$ is closed, namely $\gamma^{E_0}=\gamma^{E_1}$, then there is an alternative way to calculate the index. Let $E\in[E_0,E_1]\mapsto e^{\imath\theta_l^E}$ be continuous paths of the eigenvalues of
$U^E$ with arbitrary choice of enumeration at level crossings. Each of these paths has a winding number and clearly
\begin{equation}
\label{eq-intersec2}
\mbox{\rm I}(\gamma,{\Psi})
\;=\;
\sum_{l=1}^L
\;\mbox{Wind}
\Bigl(\;
E\in [E_0,E_1) \mapsto  e^{\imath\theta_l^E}
\;\Bigr)
\;.
\end{equation}
In particular, the r.h.s. is independent of the choice of
enumeration of the $\theta_l^E$'s at level crossings. This leads to
\begin{equation}
\label{eq-intersec3}
\mbox{\rm I}(\gamma,{\Psi})
\;=\;
\;\mbox{Wind}
\Bigl(\;
E\in [E_0,E_1) \mapsto  \det(U^E)
\;\Bigr)
\;.
\end{equation}
As $\det(U^E)=\det(\Pi(\gamma^E))/\det(\Pi([\Psi]_\sim))$ it follows that the index $\mbox{\rm I}(\gamma,{\Psi})$ can also be calculated as the winding number of $E\in [E_0,E_1) \mapsto\det(\Pi(\gamma^E))$ and is hence independent of $\Psi$ for a closed path. In this case we therefore simply write $\mbox{\rm I}(\gamma)$. Otherwise stated, $\mbox{\rm I}(\gamma)$ is the pairing of $\gamma$ with a cocycle in the integer cohomology called the Arnold cocycle \cite{Arn}. The winding number can be calculated using a lift $E\in [E_0,E_1) \mapsto\mbox{\rm Lift}(\det(\Pi(\gamma^E)))\in\RM$ obtained with the multi-branched function $z\mapsto \frac{1}{2\pi\imath}\log(z)$. Let us now define also for a non-closed path its winding integral $\mbox{\rm W}(\gamma)\in\RM$ by
$$
\mbox{\rm W}(\gamma)
\;=\;
\mbox{\rm Lift}\bigl(\det(\Pi(\gamma^{E_1}))\bigr)
\,-\,
\mbox{\rm Lift}\bigl(\det(\Pi(\gamma^{E_0}))\bigr)
\;.
$$
This is independent of the choice of the lift. In case that the path is differentiable one has
\begin{equation}
\label{eq-winding}
\mbox{\rm W}(\gamma)
\;=\;
\;
\frac{1}{2\pi\imath}\;\int^{E_1}_{E_0}dE\;\partial_E
\log\bigl(\det(\Pi(\gamma^E))\bigr)
\;. 
\end{equation}
For a closed path, $\mbox{\rm W}(\gamma)=\mbox{\rm I}(\gamma)$. Let us point out that all these definitions also work in the case of paths in the real Lagrangian Grassmannian $\LM_L^\RM$ which is more frequently studied in the literature. Then the unitaries $U^E$ are symmetric, but this does not change the above spectral flow picture. Similarly, other symmetries than the complex conjugation can be implemented without alternating the definition of the index. Another example is the quaternion symmetry \cite{SB}. Let us collect a few basic properties of the index.

\begin{proposi}
\label{prop-index} For $\Tt\in\mbox{\rm HS}(2L,\CM)$ and $\gamma$ as above, set $\Tt\,\gamma=(\Tt\,\gamma^E)_{E\in[E_0,E_1)}$. 

\vspace{.1cm}

\noindent {\rm (i)} 
Let the path $\gamma+\gamma'$ denote the concatenation with a second path $\gamma'=(\gamma^E)_{[E_1,E_2)}$. Then
$$
\mbox{\rm I}(\gamma+\gamma',{\Psi})
\;=\;
\mbox{\rm I}(\gamma,{\Psi}) + \mbox{\rm I}(\gamma',{\Psi})
\;.
$$

\vspace{.1cm}

\noindent {\rm (ii)}  Given a second path $\gamma'=({\gamma'}^E)_{[E_0,E_1)}$ and Lagrangian frame $\Psi'$, one has {\rm (}with notation {\rm \eqref{eq-Phisum})}
$$
\mbox{\rm I}(\gamma\,\widehat{\oplus}\,\gamma',{\Psi}\,\widehat{\oplus}\,\Psi')
\;=\;
\mbox{\rm I}(\gamma,{\Psi})\, +\, \mbox{\rm I}(\gamma',\Psi')
\;.
$$

\vspace{.1cm}

\noindent {\rm (iii)} One has
$\mbox{\rm I}(\Tt\,\gamma,\Tt\,{\Psi}) =\mbox{\rm I}(\gamma,\Psi)$.

\vspace{.1cm}

\noindent {\rm (iv)} For any Lagrangian $\Psi'$,
$$
\left|\,
\mbox{\rm I}(\gamma,{\Psi}) - \mbox{\rm I}(\gamma,\Psi')\,\right|
\;\leq\;L\;,
\qquad
\left|\,
\mbox{\rm I}(\Tt\,\gamma,{\Psi}) - \mbox{\rm I}(\gamma,\Psi)\,\right|
\;\leq\;L\;.
\qquad
\left|\,\mbox{\rm I}(\gamma,{\Psi}) -\mbox{\rm W}(\gamma)\,\right|\;\leq \;L
\;.
$$

\vspace{.1cm}

\noindent {\rm (v)} 
For a closed path $\gamma$, the index $\mbox{\rm I}(\gamma,{\Psi})$ is independent of $\Psi$ and $\mbox{\rm I}(\Tt\,\gamma)=\mbox{\rm I}(\gamma)=\mbox{\rm W}(\gamma)$.

\end{proposi}

\noindent {\bf Proof.} Items (i) through (iii) follow immediately from the definition. Using Proposition~\ref{prop-diffeo}, one can check that the compact subgroup $G=\mbox{\rm HS}(2L,\CM)\cap\mbox{\rm U}(2L)$ acts transitively on $\LM_L$. Hence there is an $\Tt\in G$ such that $\Psi'=\Tt\Psi$. Then there exists $\Mm\in G$ such that $\Rr=\Mm\Tt\Mm^*$ is a rotation matrix by an angle $\eta$, namely $\Cc\Rr\Cc^*=\mbox{\rm diag}(e^{\imath\eta},e^{-\imath\eta})$. Now using (iii),
$$
\mbox{\rm I}(\gamma,{\Psi}) - \mbox{\rm I}(\gamma,\Psi')
\;=\;
\mbox{\rm I}(\Mm\gamma,\Mm{\Psi}) - \mbox{\rm I}(\gamma,\Rr\Mm\Psi)
\;.
$$
But $\Pi(\Rr\Mm\Psi)=e^{2\imath\eta}\Pi(\Mm\Psi)$ so that all $L$ eigenvalues of $\Pi(\Mm\Psi)$ are shifted by $e^{2\imath\eta}$. By \eqref{eq-intersec3} this implies that the index difference can be at most $L$ . Because $\mbox{\rm I}(\Tt\,\gamma,{\Psi})=\mbox{\rm I}(\gamma,\Tt^{-1}{\Psi})$ the second estimate of (iv) is equivalent to the first one. The third one follows either from first two or directly from the definitions. In (v) the independence of $\Psi$ was already proved above and it implies the second claim.
\hfill $\Box$

\vspace{.2cm}

Next we consider a path $\Gamma=(\Tt^E)_{E\in[E_0,E_1)}$ in the group $\mbox{\rm HS}(2L,\CM)$. Given a second such path  $\Gamma'=({\Tt'}^E)_{E\in[E_0,E_1)}$, one has $\Gamma\,\Gamma'=(\Tt^E{\Tt'}^E)_{E\in[E_0,E_1)}$. Similarly the inverse path is $\Gamma^{-1}=((\Tt^E)^{-1})_{E\in[E_0,E_1)}$. Combined with a Lagrangian $\Phi$ as base point or a path $\gamma=(\gamma^E)_{E\in[E_0,E_1)}$  in $\LM_L$, each such path $\Gamma$ induces paths in the Lagrangian Grassmannian $\LM_L$:
$$
\Gamma\,\Phi
\;=\;
(\Tt^E\,[\Phi]_\sim)_{E\in[E_0,E_1)}
\;,
\qquad
\Gamma\,\gamma
\;=\;
(\Tt^E\,\gamma^E)_{E\in[E_0,E_1)}
\;.
$$
Inversely, given a path $\gamma=(\gamma^E)_{E\in[E_0,E_1)}$  in $\LM_L$, there are many different ways to associate paths in $\mbox{\rm HS}(2L,\CM)$. One family of such paths is somewhat singled out as paths in $\mbox{\rm HS}(2L,\CM)\cap\mbox{\rm U}(2L)$: choose a representative $\gamma^E=[\Phi^E]_\sim$ satisfying $(\Phi^E)^*\Phi^E=\one_L$ and then set $\Tt^E=(\Phi^E,\Jj\Phi^E)$.

\begin{proposi}
\label{prop-index2} For Lagrangian $\Phi$ and $\gamma$, $\Gamma$, $\Gamma'$ as above one has:

\vspace{.1cm}

\noindent {\rm (i)} ${\mbox{\rm I}}(\Gamma\Phi,\Psi)=
-\,{\mbox{\rm I}}(\Gamma^{-1}\Psi,\Phi)$.

\vspace{.1cm}

\noindent {\rm (ii)} For any Lagrangian $\Phi'$, 
$$
\left|\,
{\mbox{\rm I}}(\Gamma\,\Phi,\Psi)\,-\,
{\mbox{\rm I}}(\Gamma\,\Phi',\Psi)\,\right|
\;\leq\; L\;.
$$

\vspace{.1cm}

\noindent {\rm (iii)} For any Lagrangian $\Psi'$ and $\Psi''$, 
$$
\left|\,
{\mbox{\rm I}}(\Gamma\,\gamma,\Psi)\,-\,
{\mbox{\rm I}}(\Gamma\,\Phi,\Psi')\,-\,
{\mbox{\rm I}}({\gamma},\Psi'')\,\right|
\;\leq\; 5\,L\;.
$$

\vspace{.1cm}

\noindent {\rm (iv)} 
For closed paths $\gamma$ and $\Gamma$ and any $\Phi$, 
$$
{\mbox{\rm I}}(\Gamma\,\gamma)\;=\;
{\mbox{\rm I}}(\Gamma\,\Phi)\,+\,
{\mbox{\rm I}}({\gamma})
\;.
$$

\vspace{.1cm}

\noindent {\rm (v)} 
For a closed path $\Gamma$, ${\mbox{\rm I}}(\Gamma\,\Phi)$ is independent of $\Phi$. 

\end{proposi}

\noindent {\bf Proof.}  (i) The index ${\mbox{\rm I}}(\Gamma\Phi,\Psi)$ is defined in \eqref{eq-intersec} as a sum over $E$ with non-trivial intersections $\Tt^E\Phi\,\CM^L\cap\Psi\,\CM^L$. Precisely for those $E$ the intersection $(\Tt^E)^{-1}\Psi\,\CM^L\cap\Phi\,\CM^L$ is non-trivial. Moreover, the dimensions of these intersections coincide and a bit of thought shows that the signatures of the corresponding paths (as $E$ varies) have reversed signs. Now by (i),
${\mbox{\rm I}}(\Gamma\,\Phi,\Psi)-
{\mbox{\rm I}}(\Gamma\,\Phi',\Psi)={\mbox{\rm I}}(\Gamma^{-1}\Psi,\Phi')-
{\mbox{\rm I}}(\Gamma^{-1}\Psi,\Phi)$, so Proposition~\ref{prop-index}(iv) implies (ii). In order to prove (iii) we will check that
\begin{equation}
\label{eq-Windest}
\left|\,
{\mbox{\rm W}}(\Gamma\,\gamma)\,-\,
{\mbox{\rm W}}({\gamma})\,-\,
{\mbox{\rm W}}(\Gamma\,\Phi)\,\right|
\;\leq\; 2\,L\;.
\end{equation}
Then (iii) follows when combining this with Proposition~\ref{prop-index}(iv). For the proof of \eqref{eq-Windest}, denote $V=\Pi([\Phi]_\sim)$, $U=\Pi(\Gamma^E)$ and the entries of $\Cc\Tt^E\Cc^*$ as in \eqref{eq-CTCcoef}, that is, we drop the energy dependence throughout. Then, using that $A$ and $D$ are invertible ({\it cf.} proof of Proposition~\ref{prop-TtoU0}),
\begin{eqnarray*}
\det(\Pi(\Tt^E{\gamma^E}))
& = &
\det\bigl((AU+B)(CU+D)\bigr)
\\
& = &
\det(U)\,\det\bigl((AV+B)(CV+D)\bigr)\;
\frac{\det(\one+A^{-1}BU)\det(\one+D^{-1}CV)}{
\det(\one+A^{-1}BV)\det(\one+D^{-1}CU)}
\;.
\end{eqnarray*}
Now let us take the winding integral. The l.h.s. gives ${\mbox{\rm W}}(\Gamma\,\gamma)$, while $\det(U)$ and the second factor on the r.h.s. give  ${\mbox{\rm W}}({\gamma})$ and ${\mbox{\rm W}}(\Gamma\,\Phi)$. Further $\|A^{-1}B\|< 1$ because $A^*A=\one+B^*B$, so that $\log\det(\one+A^{-1}BU)=\Tr\log(\one+A^{-1}BU)$ only needs one branch of the logarithm and hence the winding integral is bounded by $\frac{L}{2}$. Similarly $\|D^{-1}C\|<1$ and thus the other three factors give also at most a contribution $\frac{L}{2}$. This implies \eqref{eq-Windest}. Considering closed paths, this argument also shows (iv). Item (v) is clear from (iv).
\hfill $\Box$

\vspace{.2cm}

We are now interested in counting the number of passages of the eigenvalues of $\Tt^E$ through $e^{\imath k}$ weighted by orientation. In view of the discussions in Section~\ref{sec-symplectic}, there is very natural way to do this: the graph of $\Tt^E$ is again a path in the Lagrangian Grassmannian $\LM_{2L}$ and the singular cycle associated to $\widehat{\Psi}_k\in\LM_{2L}$ defined in \eqref{eq-Psik} locates the matrices $\Tt^E$ with eigenvalue $e^{\imath k}$. An index of $\Gamma$ can now be defined by just applying the previous definition to the graph. Hence, associated to $\Gamma$ let us define $\widehat{\Gamma}=\bigl(\widehat{\Tt}^E\bigr)_{E\in[E_0,E_1)}$ where as above $\widehat{\Tt}^E=\one_L\widehat{\oplus}{\Tt}^E$ and then
\begin{equation}
\label{eq-CZdef}
\widehat{\mbox{\rm I}}(\Gamma,e^{\imath k})
\;=\;
{\mbox{\rm I}}\bigl(
\widehat{\Gamma}\,
[\widehat{\Psi}_0]_\sim\,,
\widehat{\Psi}_k\bigr)
\;.
\end{equation}
For a closed path $\Gamma$, the index $\widehat{\mbox{\rm I}}(\Gamma,e^{\imath k})$ is again independent of $e^{\imath k}$ and denoted by $\widehat{\mbox{\rm I}}(\Gamma)$. Several of the other properties of Proposition~\ref{prop-index} transpose immediately to $\widehat{\mbox{\rm I}}$, others are recollected in the next proposition.

\begin{proposi}
\label{prop-CZindex}  Let $\Phi$, $\Psi$, $\gamma$, $\Gamma$ and $\Gamma'$ be as above.

\vspace{.1cm}

\noindent {\rm (i)} 
For a closed path $\Gamma$, one has $\widehat{\mbox{\rm I}}(\Gamma)=
{\mbox{\rm I}}(\Gamma\,\Phi) $ independently of $\Phi$.

\vspace{.1cm}

\noindent {\rm (ii)}  For closed paths $\gamma$ and $\Gamma$, one has ${\mbox{\rm I}}(\Gamma\,\gamma)=
\widehat{\mbox{\rm I}}(\Gamma)+{\mbox{\rm I}}({\gamma})$.

\vspace{.1cm}

\noindent {\rm (iii)} For closed paths $\Gamma$ and $\Gamma'$, one has $\widehat{\mbox{\rm I}}(\Gamma\,\Gamma')=
\widehat{\mbox{\rm I}}(\Gamma)+\widehat{\mbox{\rm I}}(\Gamma')$.

\vspace{.1cm}

\noindent {\rm (iv)} One has
$\bigl|\,\widehat{\mbox{\rm I}}(\Gamma,e^{\imath k})-
{\mbox{\rm I}}(\Gamma\,\Phi,\Psi)\,\bigr|\leq2\,L\,.$

\vspace{.1cm}

\noindent {\rm (v)} One has $\bigl|\,
{\mbox{\rm I}}(\Gamma\gamma,\Psi)-
\widehat{\mbox{\rm I}}(\Gamma,e^{\imath k})-
{\mbox{\rm I}}({\gamma},\Psi)\,\bigr|
\leq 7\,L\;.$

\end{proposi}

\noindent {\bf Proof.} Because the path on the r.h.s. of \eqref{eq-CZdef} is closed, Proposition~\ref{prop-index}(v) one can replace $\widehat{\Psi}_k$ in \eqref{eq-CZdef} by $\Psi\widehat{\oplus}\Psi'$. Then apply Proposition~\ref{prop-index2}(i) to shift the matrices $\widehat{\Tt}^E$ on the other argument, and replace also $\widehat{\Psi}_0$ by a direct sum $\Phi\widehat{\oplus}\Phi'$. But now all objects split into symplectic sums and thus one can apply Proposition~\ref{prop-index}(ii). In the first argument there is no $E$-dependence and hence a vanishing contribution to the index, and the second gives ${\mbox{\rm I}}(\Gamma\,\Phi)$. Both (ii) and (iii) follow directly from (i) and Proposition~\ref{prop-index2}(iv). Combining Proposition~\ref{prop-index}(iv) and  Proposition~\ref{prop-index2}(ii), we get
$$
\left|\,\widehat{\mbox{\rm I}}(\Gamma,e^{\imath k})
\;-\;
{\mbox{\rm I}}\bigl(
\widehat{\Gamma}\,[\Phi'\widehat{\oplus}\Phi]_\sim,
\Psi'\widehat{\oplus}\Psi\bigr)
\,\right|
\;\leq\;
2\,L\;.
$$
Thus (iv) follows from Proposition~\ref{prop-index}(ii). Item (v) results from (iv) and Proposition~\ref{prop-index}(iii).
\hfill $\Box$

\section{Boundary conditions}
\label{sec-boundary}

Equivalent to \eqref{eq-matrix}, one can consider $H_N({\omega})$ as a tridiagonal operator $H_N$ acting on sequences 
$(\phi_n)_{n=0,\ldots,N+1}$ of vectors in $\CM^L$ as
\begin{equation}
\label{eq-jacobi}
(H_{N}\phi)_n
\;=\;
T_{n+1}\phi_{n+1}\,+\,V_n\phi_n
\,+\,T_{n}^*\phi_{n-1}
\;,
\qquad
n=1,\ldots,N\;,
\end{equation}
where $T_{N+1}=T_1$, together with the boundary conditions 
\begin{equation}
\label{eq-boundary}
\phi_{N+1}\;=\;\omega\,\phi_1\;,
\qquad
\phi_{0}\;=\;\overline{\omega}\,\phi_N\;.
\end{equation}
These boundary conditions are Dirichlet for $\omega=0$ and periodic for $\omega=e^{\imath k}$. It is useful to take a more abstract point of view on these boundary conditions. They are $2L$ linear relations between the four $L$-dimensional vectors $\phi_0$, $\phi_1$, $\phi_N$ and $\phi_{N+1}$, forcing them to lie in a particular $2L$-dimensional plane of $\CM^{4L}$. In principle, any other $2L$-dimensional plane of $\CM^{4L}$ leads when combined with \eqref{eq-jacobi} to an operator on a $NL$-dimensional Hilbert space. Such a plane can be described by a $4L\times 2L$ matrix $\widehat{\Psi}$ built out of $2L$ linearly independent vectors in $\CM^{4L}$ spanning the plane. Now a convenient way to formulate the generalized boundary condition is 
\begin{equation}
\label{eq-boundarygen}
\left(
\begin{array}{c}
\phi_0 \\
T_{N+1}\phi_{N+1} \\
T_1\phi_1
\\
\phi_N
\end{array}
\right)
\;=\;
\widehat{\Psi}\,v\;,
\end{equation}
for some vector $v\in\CM^{2L}$. This defines a linear operator $H_N^{\widehat{\Psi}}$ on $\ell^2(\{1,\ldots,N\},\CM^L)$. The matrices $T_{N+1}=T_1$ are introduced in \eqref{eq-boundarygen} for later convenience.   The special case \eqref{eq-boundary} with $\omega=e^{\imath k}$ corresponds to the choice $\widehat{\Psi}_k$ defined in \eqref{eq-Psik}. Now not all boundary conditions $\widehat{\Psi}$ lead to a self-adjoint operator $H_N^{\widehat{\Psi}}$. Indeed, given $\phi=(\phi_n)_{n=0,\ldots,N+1}$ and $\phi'=(\phi'_n)_{n=0,\ldots,N+1}$ corresponding to vectors $v$ and $v'$ in \eqref{eq-boundarygen}, one has
\begin{eqnarray*}
\langle H_N^{\widehat{\Psi}}\phi'\,|\,\phi\rangle
& = & 
\sum_{n=1}^N
(T_{n+1}\phi'_{n+1}\,+\,V_n\phi'_n
\,+\,T_{n}^*\phi'_{n-1})^*
\,\phi_n
\\
& = & 
\langle \phi'\,|\,H_N^{\widehat{\Psi}}\phi\rangle
\,+\,
(\phi'_{N+1})^*\,T^*_{N+1}\phi_N
\,-\,
(\phi'_{1})^*\,T^*_{1}\phi_0
\,+\,
(\phi'_{0})^*\,T_{1}\phi_1
\,-\,(\phi'_{N})^*\,T_{N+1}\phi_{N+1}
\\
& = &
\langle \phi'\,|\,H_N^{\widehat{\Psi}}\phi\rangle
\,-\,
(\widehat{\Psi}\,v')^*\,\widehat{\Jj}\,\widehat{\Psi}\,v\;.
\end{eqnarray*}
Now selfadjointness requires the last term to vanish for all $v$ and $v'$. Hence a selfadjoint boundary condition is given if and only if $\widehat{\Psi}^*\widehat{\Jj}\,\widehat{\Psi}=0$.
This means that the plane $[\widehat{\Psi}]_\sim$ is Lagrangian w.r.t. $\widehat{\Jj}$. By Proposition~\ref{prop-diffeo} the set of self-adjoint boundary conditions can hence be identified with $\mbox{\rm U}(2L)$.

\vspace{.2cm}

As already pointed out, the boundary condition for  $H^k_N$ is $\widehat{\Psi}_k$. The Dirichlet boundary conditions defining $H^{\mbox{\rm\tiny D}}_N$ are given by 
$\widehat{\Psi}_{\mbox{\rm\tiny D}}=\Psi_{\mbox{\rm\tiny L}}\widehat{\oplus}\Psi_{\mbox{\rm\tiny R}}$ where  $\Psi_{\mbox{\rm\tiny L}}=\Psi_{\mbox{\rm\tiny R}}=\binom{0}{\one_L}$. Other boundary conditions separating left and right boundary can readily be dealt with and are simply given by modifications of $V_1$ and $V_N$ \cite{SB}.

\section{Eigenfunctions and transfer matrices}
\label{sec-transfer}

As for a one-dimensional Jacobi matrix, 
it is useful to rewrite the Schr\"odinger
equation 
\begin{equation}
\label{eq-Schroedinger}
H_N^k\phi
\;=\;
E\,\phi\;,
\end{equation}
for a real energy $E\in\RM$
in terms of the $2L\times 2L$ transfer matrices $\Tt_n^E$ defined by
\begin{equation}
\label{eq-transfer}
\Tt_n^E
\;=\;
\left(
\begin{array}{cc}
(E\,{\bf 1}\,-\,V_n)\,T_n^{-1} & - T_n^* \\
T_n^{-1} & {\bf 0}
\end{array}
\right)
\;,
\qquad
n=1,\ldots,N
\;.
\end{equation}
For a real energy $E\in\RM$,
each transfer matrix is in the hermitian symplectic group $\mbox{\rm HS}(2L,\CM)$.
The  Schr\"odinger equation (\ref{eq-Schroedinger}) is satisfied if and only if
\begin{equation}
\label{eq-transferid}
\left(
\begin{array}{c}
T_{n+1}\phi_{n+1} \\
\phi_n
\end{array}
\right)
\;=\;
\Tt^E_n\,
\left(
\begin{array}{c}
T_{n}\phi_{n} \\
\phi_{n-1}
\end{array}
\right)
\;,
\qquad
n=1,\ldots,N
\;,
\end{equation}
and the boundary conditions (\ref{eq-boundary}) hold. Let us first focus on the case of periodic boundary conditions where $\omega=e^{\imath k}$ as this leads to the proof of Theorem~\ref{theo-osci} stated in the introduction. These boundary conditions can be written in a compact manner:
$$
\left(
\begin{array}{c}
T_{N+1}\phi_{N+1} \\
\phi_N
\end{array}
\right)
\;=\;
e^{\imath k}\;
\left(
\begin{array}{c}
T_{1}\phi_{1} \\
\phi_0
\end{array}
\right)
\;.
$$
Next write the l.h.s. using \eqref{eq-transferid}.
Setting $\Tt^E(N,1)=\Tt^E_N\cdots\Tt^E_1$, this means that $E$ is an eigenvalue of $H^k_N$ if and only if $e^{\imath k}$ is an eigenvalue of $\Tt^E(N,1)$. More precisely,
$$
\mbox{\rm multiplicity of }E\;\mbox{\rm as eigenvalue of }H^k_N
\,=\,
\mbox{\rm geometric multiplicity of }e^{\imath k}\;\mbox{\rm as eigenvalue of }\Tt^E(N,1).
$$
As $\Tt^E(N,1)$ is in the group $\mbox{\rm HS}(2L,\CM)$, one way to calculate the r.h.s. is to appeal to Proposition~\ref{prop-TtoU}. Hence let the expanded transfer matrices $\widehat{\Tt}^E_n=\one_{2L}\widehat{\oplus}\Tt^E_n\in\mbox{\rm HS}(4L,\CM)$ be defined as in \eqref{eq-expanding}. Then define Lagrangian frames by
\begin{equation}
\label{eq-transferidexpand}
\widehat{\Phi}^E_n
\;=\;
\widehat{\Tt}^E_n
\;\widehat{\Phi}^E_{n-1}
\;,
\qquad
\widehat{\Phi}^E_0
\;=\;
\widehat{\Psi}_0
\;,
\end{equation}
with $\widehat{\Psi}_0$ as in \eqref{eq-Phiexpand}.  Finally let us denote the associated unitaries by
$$
\widehat{W}^E_n
\;=\;
\widehat{\Pi}\bigl([ \widehat{\Phi}^E_n]_\sim\bigr)
\;,
\qquad
\widehat{U}^{E,k}_n
\;=\;
\left(
\begin{array}{cc}
0 & \imath\,e^{-\imath k}\,\one_{L} \\
\imath\,e^{\imath k}\,\one_{L} & 0
\end{array}
\right)
\;\widehat{W}^E_n
\;.
$$
Due to \eqref{eq-Moeb}, these unitaries can be calculated iteratively using the M\"obius transformation:
\begin{equation}
\label{eq-unitaryMoebius}
\widehat{W}^E_n
\;=\;
\widehat{\Cc}\,\widehat{\Tt}^E_n\,\widehat{\Cc}^*\cdot\widehat{W}^E_{n-1}
\;.
\end{equation}
Now Proposition~\ref{prop-TtoU} applied to $\Tt^E(N,1)\in\mbox{\rm HS}(2L,\CM)$  directly gives item (iii) of Theorem~\ref{theo-osci}. The other two statements are proved in Sections~\ref{sec-mono} and \ref{sec-Asymp}.

\vspace{.2cm}

Now let us consider the case of Dirichlet boundary conditions. This could be done in exactly the same manner using the Lagrangian frames $\widehat{\Phi}^E_n$ defined in \eqref{eq-transferidexpand}, but with the initial condition $\widehat{\Phi}^E_0=\Psi_{\mbox{\rm\tiny L}}\widehat{\oplus}\Psi_{\mbox{\rm\tiny R}}$, and then analysing its intersection of $\widehat{\Phi}^E_N$  with $\Psi_{\mbox{\rm\tiny R}}\widehat{\oplus}\Psi_{\mbox{\rm\tiny L}}$. As all objects are symplectic direct sums, it is easier to consider directly the $L$-dimensional Lagrangian frames
$$
{\Phi}^E_n
\;=\;
{\Tt}^E_n
\;{\Phi}^E_{n-1}
\;,
\qquad
{\Phi}^E_0
\;=\;
{\Psi}_{\mbox{\rm\tiny D}}
\;,
$$
where ${\Psi}_{\mbox{\rm\tiny D}}=\binom{\one_L}{0}$. By the intersection theory of Section~\ref{sec-symplectic}, the multiplicity of $E$ as eigenvalue of $H_N^{\mbox{\rm\tiny D}}$ is then equal to the dimension of the intersection of the Lagrangian frames ${\Phi}^E_N$ and $\Psi_{\mbox{\rm\tiny R}}=\binom{0}{\one_L}$, and hence equal to $1$ as eigenvalue of the unitary 
$$
U^E_N
\;=\;
\Pi([
\Psi_{\mbox{\rm\tiny R}}]_\sim)^*\,
\Pi([\Phi^E_N]_\sim)
\;=\;
-\;\Pi([\Phi^E_N]_\sim)
\;.
$$
This provides the proof of the first statement of the following theorem. The proof of the second is similar to that of Proposition~\ref{prop-derivbound} below. 

\begin{theo} 
\label{theo-osci2} {\rm \cite{SB}}
The multiplicity of $E$ as eigenvalues $H_N^{\mbox{\rm\tiny D}}$ is equal to the multiplicity of $1$ as eigenvalue of $U^E_N$.  As a function of energy $E$, the eigenvalues of ${U}^{E}_N$ rotate around the unit circle in the positive sense and with non-vanishing speed.
\end{theo}

Let us point out that in the case $L=1$ the unitaries $U^E_N$ are called Pr\"ufer phases. Hence it is natural to call them matrix Pr\"ufer phases for general $L$.

\section{Monotonicity}
\label{sec-mono}

According to the above, the spectrum of $H^k_N$ can be read off the spectrum of the path of unitaries $E\in\RM\mapsto \widehat{U}^{E,k}_N$. We next show that the eigenvalues of this unitary all rotate in the same direction, namely item (i) of Theorem~\ref{theo-osci}. The latter follows from the positivity of the operator appearing in the following proposition, because the rotation speeds of the eigenvalues of $\widehat{U}^{E,k}_N$ are given by the diagonal matrix elements of this operator w.r.t. to the basis of eigenvectors of $\widehat{U}^{E,k}_N$.

\begin{proposi}
\label{prop-derivbound}
For $E\in\RM$ and $N\geq 2$, the matrix
$$
\frac{1}{\imath}\,(\widehat{U}^{E,k}_N)^*\,\partial_E\,\widehat{U}^{E,k}_N
\;=\;
\frac{1}{\imath}\,(\widehat{U}^E_N)^*\,\partial_E\,\widehat{U}^E_N
$$
is positive semi-definite.
\end{proposi}

\noindent {\bf Proof.}  (In part, similar to \cite{SB}) Let us introduce
$\phi^E_\pm=(\,\one\;\pm\imath\one\,)\;\widehat{\Phi}^E_N$. These are
invertible $2L\times 2L$ matrices by the argument in Proposition~\ref{prop-diffeo} and one has
$\widehat{U}^E_N=\phi_-^E(\phi_+^E)^{-1}=((\phi_-^E)^{-1})^*(\phi_+^E)^*$. Now
$$
(\widehat{U}^E_N)^*\,\partial_E\,\widehat{U}^E_N
\;=\;
((\phi_+^E)^{-1})^*
\Bigl[\,
(\phi_-^E)^*\partial_E\phi_-^E \,-\,
(\phi_+^E)^*\partial_E\phi_+^E
\,\Bigr]
(\phi_+^E)^{-1}
\;.
$$
Thus it is sufficient to verify positive definiteness of
$$
\frac{1}{\imath}\;
\Bigl[\,
(\phi_-^E)^*\partial_E\phi_-^E \,-\,
(\phi_+^E)^*\partial_E\phi_+^E
\,\Bigr]
\;=\;
2\;
(\widehat{\Phi}^E_N)^*\,\widehat{\Jj}\,\partial_E \widehat{\Phi}^E_N
\;.
$$
From the product rule follows that
$$
\partial_E \widehat{\Phi}^E_N
\;=\;
\sum_{n=1}^N
\;
\left(
\prod_{l=n+1}^N\,\widehat{\Tt}^E_l
\right)
\,
\left(\partial_E \widehat{\Tt}^E_n\right)
\;
\left(
\prod_{l=1}^{n-1}\,\widehat{\Tt}^E_l
\right)
\,\widehat{\Phi}_0^E
\;.
$$
This implies that
$$
(\widehat{\Phi}^E_N)^*\,\widehat{\Jj}\,\partial_E \widehat{\Phi}^E_N
\;=\;
\sum_{n=1}^N
\;(\widehat{\Phi}_0^E)^*\,
\left(
\prod_{l=1}^{n-1}\,\widehat{\Tt}^E_l
\right)^*
\,
\bigl(\widehat{\Tt}^E_n\bigr)^*\,\widehat{\Jj}\,
\bigl(\partial_E \widehat{\Tt}^E_n\bigr)
\;
\left(
\prod_{l=1}^{n-1}\,\widehat{\Tt}^E_l
\right)
\,\widehat{\Phi}_0^E
\;.
$$
As one checks that
$$
\bigl(\widehat{\Tt}^E_n\bigr)^*\,\Jj\,
\bigl(\partial_E \widehat{\Tt}^E_n\bigr)
\;=\;
\left(
\begin{array}{cccc}
0 &  0 & 0 & 0 \\
0 & (T_nT_n^*)^{-1} & {0} & 0 \\
0 &  0 & 0 & 0 \\ 
0 &  0 & 0 & 0 
\end{array}
\right)
\;,
$$
and the matrices $\widehat{\Tt}^E_n$ do not mix first and third columns and lines with the second and forth ones, it follows that
$$
(\widehat{\Phi}^E_N)^*\,\widehat{\Jj}\,\partial_E \widehat{\Phi}^E_N
\;=\;
\sum_{n=1}^N
\;
\left(
\prod_{l=1}^{n-1}\,{\Tt}^E_l
\right)^*
\,
\left(
\begin{array}{cc}
(T_nT_n^*)^{-1} & {0}  \\
0 &  0
\end{array}
\right)
\;
\left(
\prod_{l=1}^{n-1}\,{\Tt}^E_l
\right)
\;.
$$
Clearly each of the summands is positive semi-definite. 
In order to prove the strict inequality, it is sufficient that the first two
terms $n=1,2$ give a strictly positive contribution. Hence 
let us verify that
$$
\bigl(\Tt^E_2\bigr)^*\;
\left(
\begin{array}{cc}
(T_1T_1^*)^{-1} & { 0} \\
{0} & { 0}
\end{array}
\right)
\;\Tt^E_2
\;+\;
\left(
\begin{array}{cc}
(T_2T_2^*)^{-1} & {0} \\
{0} & {0}
\end{array}
\right)
\;>\;
0
\;.
$$
For this purpose let us show that the kernel of the matrix on the l.h.s. is
empty. As $\bigl((\Tt^E_2)^*\bigr)^{-1}=-\Jj\Tt^E_2\Jj$, we thus have to show
that a vector $\left(\begin{array}{c} v \\ w \end{array}
\right)\in\CM^{2L}$ satisfying
$$
-\;\Jj\;
\left(
\begin{array}{cc}
(T_1T_1^*)^{-1} & {0} \\
{0} & {0}
\end{array}
\right)
\;\Tt^E_2
\;
\left(\begin{array}{c} v \\ w \end{array}
\right)
\;=\;
\Tt^E_2\;\Jj\;
\left(
\begin{array}{cc}
(T_2T_2^*)^{-1} & {0} \\
{0} & {0}
\end{array}
\right)
\;\left(\begin{array}{c} v \\ w \end{array}
\right)
\;,
$$
actually vanishes. Carrying out the matrix multiplications, one readily checks
that this is the case.
\hfill $\Box$

\vspace{.2cm}

The matrix $\widehat{U}^E_N$ can be calculated using the iterative M\"obius transformation, see \eqref{eq-unitaryMoebius}. In parallel, the positive matrices $\frac{1}{\imath}(\widehat{U}^E_n)^*\partial_E\widehat{U}^E_n$ can be calculated iteratively. The corresponding equations can be readily written out, but as they are a bit lengthy and not used here we refrain from doing so.

\vspace{.2cm}

%
%
%

\section{Asymptotics}
\label{sec-Asymp}

The following result proves Theorem~\ref{theo-osci}(ii).

\begin{proposi}
\label{prop-asymptotics}
The asymptotics of the unitaries are
$$
\lim_{E\to\pm\infty}\;\widehat{U}^{E,k}_N\;=\;
\left(
\begin{array}{cc}
0 & \imath\,e^{-\imath k}\,\one_{L} \\
\imath\,e^{\imath k}\,\one_{L} & 0
\end{array}
\right)
\;.
$$
\end{proposi}

\noindent {\bf Proof.}  It is sufficient to check $\lim_{E\to\pm\infty}\widehat{W}^{E}_N=\one_{2L}$. 
We shall use $\widehat{W}^{E}_N=\widehat{\Cc}\,\widehat{\Tt}^E(N,1)\widehat{\Cc}^*\cdot \widehat{\Pi}([\widehat{\Phi}_0]_\sim)$ and the formula given in Proposition~\ref{prop-TtoU0}. Hence let $A,B,C,D$ be the entries of ${\Tt}^E(N,1)$ and $A',B',C',D'$ those of $\Cc\,{\Tt}^E(N,1)\Cc^*$. Their link can be read off
from \eqref{eq-linkC}. One has $A=E^N\prod_{n=1}^N T^{-1}+\Oo(E^{N-1})$ while $B,C,D$ are all of order $\Oo(E^{N-1})$. Hence $A',B',C',D'$ are all equal to $\frac{1}{2}\,E^N\prod_{n=1}^N T^{-1}$ up to terms of order $\Oo(E^{N-1})$. Hence by Proposition~\ref{prop-TtoU0}
$$
\widehat{W}^{E}_N
\;=\;
\left(
\begin{array}{cc}
\one_L+\Oo(E^{-1}) & \Oo(E^{-N}) \\
R & \one_L+\Oo(E^{-1}) 
\end{array}
\right)
\;.
$$
Even though not much can be deduced directly about the lower left entry $R$, unitarity of $\widehat{W}^{E}_N$ imposes $R=\Oo(E^{-1})$.
\hfill $\Box$

\vspace{.2cm}

Proposition~\ref{prop-asymptotics} implies that the path $E\in\RM\mapsto \widehat{U}^{E,k}_N$ can be closed using the one-point compactification of $\RM$. As such it can also be seen as a closed path $\gamma=(\gamma^E)_{E\in\RM}$ in the Lagrangian Grassmannian $\LM_{2L}$ if we set $\gamma^E=[\widehat{\Tt}^E(N,1)\,\widehat{\Psi}_k]_\sim$. By Propositions~\ref{prop-index2}(i) and \ref{prop-index}(ii) its index is equal to the index of the path $\gamma'=(\gamma'^E)_{E\in\RM}$ in $\LM_L$ with $\gamma'={\Tt}^E(N,1)\Phi$ for any choice of $\Phi\in\LM_L$. The index can readily be calculated by a homotopy argument and is equal to $NL$ \cite{SB}. Furthermore, by the results above, this index of $\gamma$ is equal to the total number of eigenvalues of $H^k_N$ and thus equal to $NL$. It is reassuring that this can also be calculated independently.

\section{Eigenvalues of linear Hamiltonian systems}
\label{sec-HamiltonianSys}

In this section, we briefly illustrate how the techniques of Sections~\ref{sec-symplectic} and \ref{sec-index}  can be used to study the solutions of linear Hamiltonian systems of the form:
\begin{equation}
\label{eq-HamSys}
\bigl(\Jj\,\partial_x\,+\,\Vv(x)\bigr)\,\Phi(x)\;=\;E\;\Pp(x)\,\Phi(x)\;,
\qquad
\Phi\in H^1((0,1),\CM^{2L})
\;,
\end{equation}
where $\Vv(x)=\Vv(x)^*$ and $\Pp(x)\geq 0$ are continuous functions on the finite interval $[0,1]$  into the hermitian and respectively non-negative matrices of size $2L\times 2L$. Again $E\in\RM$ is a spectral parameter. Let us focus on two examples. If $\Pp(x)=\one_{2L}$, then \eqref{eq-HamSys} is the Schr\"odinger equation for the quasi-one-dimensional Dirac operator $H=\Jj\partial+\Vv$. If $\Pp(x)=\binom{\one_L\,0}{0\;\;\;0}$, then \eqref{eq-HamSys} is a rewriting of the Schr\"odinger equation $h\phi=E\phi$ with $\phi\in H^2((0,1),\CM^{L})$ for a matrix-valued Sturm-Liouville operator:
$$
h\;=\;
-\partial_x\bigl(p\partial_x+q\bigr)\;+\;q^*\partial_x\,+\,v
\;,
$$
where $p$, $q$ and $v=v^*$ are continuous functions on $[0,1]$ into the $L\times L$ matrices and $p$ is a continuously differentiable and positive with a uniform lower bound $p\geq c\,\one_L$, $c>0$. Indeed \cite{Bot}, the equivalence of  $h\phi=E\phi$ with \eqref{eq-HamSys} follows by setting
\begin{equation}
\label{eq-identify}
\Phi\;=\;
\left(\begin{array}{c}
\phi \\
(p\partial_x+q)\phi  
\end{array}
\right)
\;,
\qquad
\Vv\;=\;
\left(\begin{array}{cc}
v-q^*p^{-1}q &q^*p^{-1} \\
p^{-1}q & -p^{-1}
\end{array}
\right)
\;.
\end{equation}
Now a solution $\Phi$ of \eqref{eq-HamSys} is continuous and has left and right limits $\Phi(0)$ and $\Phi(1)$ which intervene in
$$
\langle \Phi\,|\,H\,\Phi'\rangle\,-\,\langle H\,\Phi\,|\,\Phi'\rangle
\;=\;
\Phi(1)^*\Jj\Phi'(1)\,-\,\Phi(0)^*\Jj\Phi'(0)\;,
\qquad
\Phi,\Phi'\in H^1((0,1),\CM^{2L})
\;,
$$
with scalar product on the l.h.s. in $L^2((0,1),\CM^{2L})$, as well as in
$$
\langle \phi\,|\,h\,\phi'\rangle\,-\,\langle h\,\phi\,|\,\phi'\rangle
\;=\;
\Phi(1)^*\Jj\Phi'(1)\,-\,\Phi(0)^*\Jj\Phi'(0)\;,
\qquad
\phi,\phi'\in H^2((0,1),\CM^{L})
\;,
$$
where the scalar product on the l.h.s. is now in $L^2((0,1),\CM^{L})$ and $\Phi$, $\Phi'$ on the r.h.s. are given by \eqref{eq-identify}. Note that the r.h.s. in both of these equations are the same and similar to those of the discrete case, so that also the following is completely analogous to the discrete case. Hence self-adjoint boundary conditions for $H$ and $h$ are given by
$$
\widehat{\Qq}\;\binom{\Phi(1)}{\Phi(0)}\;\in\;\widehat{\Psi}\;\CM^{2L}\;,
\qquad
\widehat{\Qq}\;=\;
\left(\begin{array}{cccc}
0 & 0 & 0 & \one_L \\
\one_L & 0 & 0 & 0 \\
0 & 0 & \one_L & 0 \\
0 & \one_L & 0 & 0
\end{array}
\right)
\;,
$$
where $\widehat{\Psi}$ is Lagrangian w.r.t. $\widehat{\Jj}$. Note that the permutation matrix $\widehat{\Qq}$ is also  underlying \eqref{eq-boundarygen}. Again the Dirichlet boundary condition is given by ${\Psi}_{\mbox{\rm\tiny D}}$ and the periodic boundary condtion by $\widehat{\Psi}_k$. Further the transfer matrix of the Jacobi matrices is replaced by the fundamental solution $\Tt^E(x)$ of \eqref{eq-HamSys}, namely given by
\begin{equation}
\label{eq-fundamental}
\partial_x \Tt^E(x)\;=\;
\Jj\,\bigl(E\,\Pp(x)\,-\,\Vv(x)\bigr)\,\Tt^E(x)\;,
\qquad
\Tt^E(0)\;=\;\one_{2L}
\;.
\end{equation}
Finally set
$$
U^E\;=\;-\,\Pi\bigl([\Tt^E(1){\Psi}_{\mbox{\rm\tiny D}}]_\sim\bigr)
\;,
\qquad
\widehat{U}^{E,k}\;=\;
\widehat{\Pi}\bigl([\widehat{\Psi}_k]_\sim\bigr)^*\;
\widehat{\Pi}\bigl([\widehat{\Tt}^E(1)\widehat{\Psi}_0]_\sim\bigr)
\;.
$$
%

\begin{theo} 
\label{theo-osci3} The number of linear independent solutions of {\rm \eqref{eq-HamSys}} at energy $E$ with Dirichlet boundary condition or respectively periodic boundary conditions is equal to the multiplicity of $1$ as eigenvalue of $U^E$ or respectively $\widehat{U}^{E,k}$.  As a function of energy $E$, the eigenvalues of ${U}^{E}$ and $\widehat{U}^{E,k}$ rotate around the unit circle in the positive sense.
\end{theo}

\noindent {\bf Proof.} The proof of the first statement parallels that of the discrete case. For the second, let us focus on $U^E$ an prove that $\frac{1}{\imath}(U^E)^*\partial_EU^E\geq 0$. By an argument similar to the one in the proof of Proposition~\ref{prop-derivbound}, it is sufficient to check that $\Tt^E(1)^*\Jj\partial_E\Tt^E(1)\geq 0$. For that purpose, let $\epsilon>0$. By \eqref{eq-fundamental},
$$
\partial_x\,\left(\Tt^E(x)^*\,\Jj\,\Tt^{E+\epsilon}(x)\right)
\;=\;
\epsilon\;\Tt^E(x)^*\,\Pp(x)\,\Tt^{E+\epsilon}(x)
\;.
$$
As $\Tt^E(1)^*\,\Jj\,\Tt^{E}(1)=\Jj=\Tt^E(0)^*\,\Jj\,\Tt^{E+\epsilon}(0)$, one thus has
\begin{eqnarray*}
\Tt^E(1)^*\Jj\partial_E\Tt^E(1)
& = & 
\lim_{\epsilon\to 0}\;
\epsilon^{-1}\,
\left(\Tt^E(1)^*\,\Jj\,\Tt^{E+\epsilon}(1)\,-\,
\Tt^E(0)^*\,\Jj\,\Tt^{E+\epsilon}(0)
\right)
\\
& = &
\int^1_0dx\;\Tt^E(x)^*\,\Pp(x)\,\Tt^E(x)
\;.
\end{eqnarray*}
Because $\Pp$ is positive, this implies the second claim.
\hfill $\Box$

\vspace{.2cm}

One of the differences with the discrete case is that $U^E$ and $\widehat{U}^{E,k}$ do in general not have asymptotics as $E\to\infty$ or $E\to-\infty$. Both limits don't exist in the case of $H$, while for $h$ the limit for $E\to -\infty$ does exist, but not for $E\to\infty$. The reason is that $H$ is neither bounded from above or below, while $h$ is bounded from below.

\vspace{.2cm}

As a final comment let us exhibit another type of positivity that is intrinsic to \eqref{eq-HamSys}, which is actually the one discovered by Lidskii \cite{Lid}.  Set
$$
U^E(x)
\;=\;
-\,\Pi\bigl([\Tt^E(x){\Psi}_{\mbox{\rm\tiny D}}]_\sim\bigr)
\;.
$$
Similar as in Theorem~\ref{theo-osci3}, this unitary allows to calculate the solutions of \eqref{eq-HamSys} on the interval $(0,x)$ with Dirichlet boundary conditions at $0$ and $x$. By a calculation similar as in the proof of Proposition~\ref{prop-derivbound}, one checks that
$$
\frac{1}{\imath}\;U^E(x)^*\,\partial_x\,U^E(x)
\;=\;
\left(\bigl(\Tt^E(x){\Psi}_{\mbox{\rm\tiny D}}\phi_+(x)\bigr)^{-1}\right)^*\,
\bigl(E\,\Pp(x)-\Vv(x)\bigr)\,
\bigl(\Tt^E(x){\Psi}_{\mbox{\rm\tiny D}}\phi_+(x)\bigr)^{-1}
\;,
$$
where $\phi_+(x)=(\one_L\;\imath\,\one_L)\Tt^E(x){\Psi}_{\mbox{\rm\tiny D}}$. Hence as long as $E\,\Pp(x)-\Vv(x)> 0$, the eigenvalues of $U^E(x)$ rotate in the positive sense as a function of $x$. In the case where the Hamiltonian system results from a Sturm-Liouville operator, this can be understood by the following quantum mechanical analogy: as $x$ grows, the particle described by $h$ becomes wider and hence the eigenvalue distances decrease so that the eigenvalues of $U^E(x)$ have already made more rotations up to $E$.



\begin{thebibliography}{99}

\bibitem[Arn]{Arn} V. L. Arnold, {\sl Characteristic class entering in 
quantization conditions}, Funct. Anal. Applic. {\bf 1}, 1-13 (1967).

\bibitem[Bot]{Bot} R. Bott, {\sl On the Iteration of Closed Geodesics and the
Sturm Intersection Theory}, Commun. Pure Appl. Math. {\bf 9}, 171-206 (1956).

\bibitem[CZ]{CZ} C.~E. Conley, E. Zehnder, {\sl Morse-type index theory for flows and periodic solutions of Hamiltonian equations}, Commun. Pure Appl. Math. {\bf 37}, 207-253 (1978).

\bibitem[FF]{FF} S. Friedland, R.~J.~Freitas, {\sl Revisiting the Siegel upper half plane I},
Lin. Alg. and its Appl. {\bf 376}, 19-44 (2004).

\bibitem[Gos]{Gos} M. de Gosson,
{\sl On the usefulness of an index due to Leray for studying the intersections of Lagrangian and symplectic paths},  J. Math. Pures Appl. {\bf 91},  598-613 (2009).

\bibitem[Kre]{Kre} M. G. Krein, {\sl Principles of the theory of} $\lambda${\sl -zones of stability of a canonical system of linear differential equations with periodic coefficients}, Memory of A.A.~Andronov, pp. 413-498, Izdat. Akad. Nauk SSSR, Moscow, 1955; English Transl. in: M. G. Krein, {\sl Topics in differential and integral equations and operator theory}, (Birkh\"auser, Boston, 1983).

\bibitem[Lid]{Lid} V.~B. Lidskii, {\sl Oscillation theorems for canonical systems of differential equations},
Dokl. Akad. Nauk SSSR {\bf 102}, 877-880 (1955).

\bibitem[Lon]{Lon}  Y. Long, {\sl Index theory for symplectic paths with application}, (Springer, Berlin, 2002)

\bibitem[Mas]{Mas} V.~P. Maslov, {\sl Theory of Perturbations and Asymptotic Methods} (Russian), (Ed. of Univ. Moscow, 1965). 
 
\bibitem[RS]{RS} J.~W. Robbin, D.~A. Salamon, {\sl The Maslov index for paths}, Topology {\bf 32},  827-844 (1993).

\bibitem[Rue]{Rue} D. Ruelle, {\sl Rotation numbers for
diffeomorphisms and flows}, Ann. Inst. Henri Poincar{\'e} {\bf 42},
109-115 (1985).

\bibitem[SB]{SB} H. Schulz-Baldes, 
{\sl Rotation numbers for Jacobi matrices with matrix entries}, Math. Phys. Electronic J. {\bf 13}, 40 pages (2007).


\end{thebibliography}
\end{document}